\documentclass[pre,aps,preprint,showpacs,groupedaddress]{revtex4}

\begin{document}
\title{Lagrangian Probability Distributions
of Turbulent Flows}
\author{R. Friedrich}
\affiliation{Institute for Theoretical Physics, University of M\"unster,
Wilhelm-Klemm-Str. 9, 48149 M\"unster, Germany} 
\date{\today}

\begin{abstract}
We outline a statistical theory of
turbulence based on the Lagrangian formulation of fluid motion.
We derive a hierarchy of evolution equations for 
Lagrangian N-point probability distributions as well as a 
functional equation for a suitably defined probability functional
which is the analog of Hopf's functional equation. 
Furthermore, we adress the derivation of  
a generalized Fokker-Plank equation for the joint 
velocity-position probability density of N fluid particles.
\end{abstract}

\pacs{47.27.Eq,02.50.Fz,05.40.+j}
\maketitle

The strategy of approaching the phenomenon of fully
developed turbulence by considering the
statistics of Lagrangian fluid particles
has a long tradition dating back to the early works of 
Taylor \cite{Taylor}, Richardson \cite{Richardson}
(for an overview, see \cite{Monin1}, \cite{Monin2}, \cite{Frisch}).
Recently, interest in the Lagrangian statistics
has been renewed by rigorous results on
passive scalar advection in the Kraichnan model \cite{Gawed}. 
Additionally, experimental
progress has opened the way 
to gain accurate data on the motion of tracer particles which allows
one to directly evaluate the statistics of the particle acceleration
\cite{Boden1}, \cite{Boden2} ,\cite{Boden3}, \cite{Lyon}. 
These experiments will also shed
light on the physics of relative dispersion of fluid particles
\cite{Ott}. 
 
A statistical formulation of the problem of turbulence starting
from the Lagrangian point of view will lead to considerable  
progress in modeling turbulent flows by Lagrangian
pdf method \cite{Pope1}, \cite{Pope3} \cite{Sawford}. 
This method can successfully 
deal with passive scalar transport, turbulent flows involving
chemical reactions or combustion \cite{Pope2}. In this approach,
which origionally dates back to Oboukhov \cite{Oboukhov}, 
Fokker-Planck equations are used to model the statistical
behaviour of fluid particles.
Although this is an appealing approach, only few
efforts have been made to relate the Lagrangian turbulence
statistics to diffusion processes by a direct
investigation of the Navier-Stokes equation. An exception
is the work of Heppe \cite{Heppe}, who uses a projector-formalism to
obtain a generalized diffusion equation for the joint
velocity-position probability distribution of one particle.

In the following we shall present a
hierarchy of evolution equations for N-point
probability distributions describing the behaviour of
N particles in a turbulent flow. The hierarchy is in close analogy to
the one presented by Lundgren \cite{Lundgren} and Ulinich and Ljubimov
\cite{Ulinich} for the probability distributions of the Eulerian 
velocity field (see \cite{Monin2}). 
Additionally we derive a functional equation
for a suitably defined probability functional, which is
the analog of Hopf's functional equation \cite{Hopf} and, therefore,
is a concise formulation of the problem of turbulence
in the Lagrangian formulation.   

Then we shall adress the question whether a
type of diffusion process in the sense of the random force
method advocated by Novikov \cite{Novikov}
can approximate the motion of N particles. 
We shall derive a generalized
Fokker-Planck equation involving memory terms, which determines the
evolution of the N-point probability distribution. However, the
drift and diffusion terms of the
generalized Fokker-Planck equation are expressed in terms
of conditional probabilities of higher order such that the problem 
remains unclosed. 
In a following paper 
we shall adress closure approximations.

\section{Formal Lagrangian description}

In the present section we shall introduce a formal Lagrangian 
description of fluid flow, which will be suitable for formulating
evolution equations for statistical quantities.
We consider the Navier-Stokes equation
for an incompressible Eulerian velocity field ${\bf u}({\bf x},t)$:
\begin{equation}
\frac{\partial }{\partial t} {\bf u}({\bf x},t)+{\bf u}({\bf x},t)
\cdot \nabla {\bf u}({\bf x},t)=-\nabla p({\bf x},t)+\nu \Delta
{\bf u}({\bf x},t) \qquad .
\end{equation}
In order to obtain a closed evolution equation one
has to express the pressure as a functional of
the velocity field. As is well known the pressure is governed by the 
Poisson equation
\begin{equation}
\Delta p({\bf x},t)=-\nabla \cdot [{\bf u}({\bf x},t)\cdot
\nabla {\bf u}({\bf x},t)] \qquad .
\end{equation}
In the case of a finite fluid volume V appropriate boundary conditions
have to be formulated. Let us consider
the case with a prescribed normal component of the pressure gradient.
The solution of this von Neumann boundary value problem reads 
\begin{eqnarray}\label{druck0}
p({\bf x},t)&=&\frac{1}{4\pi}
\int_V d{\bf x}
G(|{\bf x}-{\bf x}'|)
\nabla \cdot [{\bf u}({\bf x}',t)\cdot \nabla 
{\bf u}({\bf x}',t)]
\nonumber \\
&+& 
\frac{1}{4\pi}
\int_{\delta V}
G({\bf x}-{\bf x}') \nabla
p({\bf x}',t) \cdot {\bf dA}
\end{eqnarray}
Here, $G({\bf x}-{\bf x}')$ denotes the Green's function
\begin{eqnarray}
\Delta G({\bf x}-{\bf x}') &=&-4\pi \delta({\bf x}-{\bf x}')
\nonumber \\
{\bf n} \cdot \nabla_{\bf x'} G({\bf x}-{\bf x}') &=&- \frac{4 \pi}{S}
\qquad {\bf x}' \in \delta V \qquad .
\end{eqnarray}
S is the area of the surface $\delta V$ enclosing the fluid.
The boundary condition is incorporated into the definition of
the Green's function.
  
In the case of an infinitely extended fluid volume the 
boundary term vanishes and the Green's function is given by
\begin{equation}
G({\bf x},{\bf x}')= \frac{1}{|{\bf x}-{\bf
x}'|} \qquad .
\end{equation}

Now we can state the Navier-Stokes equation:
\begin{eqnarray}
\frac{\partial}{\partial t} {\bf u}({\bf x},t) &+& {\bf u}({\bf x},t)
\cdot \nabla {\bf u}({\bf x},t) 
\\
&=&
-\int_V d{\bf x}' 
\Gamma({\bf x},{\bf x}'): {\bf u}({\bf x}',t):{\bf u}({\bf x}',t)
+ {\bf F}({\bf x},t)
\nonumber \\
&+&\nu 
\int d{\bf x}' L({\bf x},{\bf x}') {\bf u}({\bf x}',t)
\qquad .
\nonumber 
\end{eqnarray}
In order to obtain a convenient formulation for the subsequent statistical
treatment we have introduced the following notation for the pressure 
and the viscous terms, respectively:
\begin{eqnarray}
[\Gamma({\bf x},{\bf x}'): 
{\bf u}({\bf x}',t):{\bf u}({\bf x}',t)]_{\alpha} &=&
\frac{\partial^3}{\partial x_\alpha\partial x_\beta
\partial x_\gamma} \frac{1}{4\pi} G({\bf x}-{\bf x}')
u_{\beta}({\bf x}',t) u_{\gamma}({\bf x}',t)
\nonumber \\
{\bf F}({\bf x},t) &=&
-\frac{1}{4\pi}
\int_{\delta V}
\nabla_{\bf x} G({\bf x}-{\bf x}'){\bf n} \cdot \nabla
p({\bf x}',t)] \cdot {\bf dA}
\nonumber \\
 L({\bf x},{\bf x}')&=&\Delta_{\bf x} \delta({\bf x}-{\bf x}')
\qquad .
\end{eqnarray} 
The quantity $L({\bf x},{\bf x}')$ is a generalized function
and is defined in a formal sense.
The term ${\bf F}({\bf x},t)$ is due to boundary conditions,
c.f. eq. (\ref{druck0}).
Note, that we have assumed that the normal pressure gradient is prescribed
at the boundary.

Now we turn to a Lagrangian formulation of the equation of fluid
motion.
To this end we consider a Lagrangian path ${\bf X}(t,{\bf y})$
of a fluid particle, which initially was located at 
${\bf X}(t_0,{\bf y})={\bf y}$. The velocity of the particle
is given in terms of the Eulerian velocity field 
${\bf u}({\bf X}(t,{\bf y}),t)$, whereas the Navier-Stokes equation
takes the form 
\begin{eqnarray}
\frac{d}{dt}{\bf X}(t,{\bf y})
&=&{\bf u}({\bf X}(t,{\bf y}),t)
\\
\frac{d}{dt}{\bf u}({\bf X}(t,{\bf y}),t) &=&
-\int d{\bf x}'\Gamma({\bf X}(t,{\bf y}),{\bf x}')
:{\bf u}({\bf x}',t)
:{\bf u}({\bf x}',t)
\nonumber \\
&+& F({\bf X}(t,{\bf y}),t)+\nu 
\int d{\bf x}' L({\bf X}(t,{\bf y}),{\bf x}') 
{\bf u}({\bf x}',t)
\qquad .
\nonumber 
\end{eqnarray}
For the evaluation of the integrals we perform a coordinate 
transformation
\begin{equation}
{\bf x}'={\bf X}(t,{\bf y}') \qquad .
\end{equation}
Due to incompressibility, the Jacobian equals unity: 
\begin{equation}\label{Jac}
Det[\frac{{\partial {\bf X}_\alpha(t,{\bf y})}}{{\partial y_\beta}}]=1
\qquad . 
\end{equation}
Now we define the Lagrangian velocity ${\bf U}(t,{\bf y})$ according to
\begin{equation}
{\bf U}(t,{\bf y})={\bf u}({\bf X}(t,{\bf y}),t) \qquad .
\end{equation}
As a result we end up with the following Lagrangian formulation of the
basic fluid dynamics equation:
\begin{eqnarray}\label{lagra}
\frac{d}{dt}{\bf X}(t,{\bf y})
&=&{\bf U}(t,{\bf y})
\\
\frac{d}{dt}{\bf U}(t,{\bf y})) &=&
-\int d{\bf y}'\Gamma[{\bf X}(t,{\bf y}),{\bf X}(t,{\bf y}')]
:{\bf U}(t,{\bf y}')
:{\bf U}(t,{\bf y}')
\nonumber \\
&+& F({\bf X}(t,{\bf y}),t)+\nu 
\int d{\bf y}' L[{\bf X}(t,{\bf y}),{\bf X}(t,{\bf y}')] 
{\bf U}(t,{\bf y}')
\qquad .
\nonumber 
\end{eqnarray}
We have obtained a representation of the acceleration of a Lagrangian
particle, although this expression is formal due to the appearance of
the generalized function $L({\bf x},{\bf x}')$.
Nevertheless, we shall find that one recovers meaningful and
well defined expressions when one proceeds to a statistical
formulation.

\section{Eulerian and Lagrangian probability distributions}

The purpose of the present section is to develop a statistical
description of the fluid motion. The central quantities will be
N-point position-velocity probability densities and, as the most
general quantity, a velocity-position probability functional.
  
We start by defining the N-point Lagrangian velocity-position distribution
function
\begin{eqnarray}\label{lg}
\lefteqn{
f^N(\lbrace{\bf u}_j,{\bf x}_j,{\bf y}_j\rbrace;t)
=}
\nonumber \\
&&
<\delta[{\bf x}_1-{\bf X}(t,{\bf y}_1)]
\delta[{\bf u}_1-{\bf U}(t,{\bf y}_1)]
....
\delta[{\bf x}_N-{\bf X}(t,{\bf y}_N)]
\delta[{\bf u}_N-{\bf U}(t,{\bf y}_N)]> \qquad .
\end{eqnarray}
This distribution function allows one to statistically
characterize the behaviour of a swarm of N fluid particles.
The brackets indicate averaging with respect to a suitably defined
statistical ensemble. Since we have ${\bf X}(t=0,{\bf y})={\bf y}$
the initial condition at $t=0$ reads:
\begin{eqnarray}\label{lg1}
\lefteqn{
f^N(\lbrace{\bf u}_j,{\bf x}_j,{\bf y}_j\rbrace;t=0)
=}
\nonumber \\
&&
\delta({\bf x}_1-{\bf y}_1)....
\delta({\bf x}_N-{\bf y}_N)
g^N(\lbrace{\bf u}_j,{\bf y}_j\rbrace) \qquad ,
\end{eqnarray}
where     
$g^N(\lbrace{\bf u}_j,{\bf y}_j\rbrace)$ is the joint probability 
distribution
for the N velocities of the fluid particles at initial time $t=0$. 
In the stationary case, 
$g^N(\lbrace{\bf u}_j,{\bf y}_j\rbrace)$ is the Eulerian probability
distribution defined below. 

The N-point probability distribution for the particle locations,
$p^N(\lbrace {\bf x}_j,{\bf y}_j\rbrace;t)$ is obtained by 
integration with respect to the velocities ${\bf u}_j$ and is the
central quantity in the theory of dispersion of particles in turbulence.
Alternatively, we may integrate eq. (\ref{lg}) over the spatial variables and obtain
the pure velocity probability distributions of the particles.

It is convenient to additionally define the corresponding Eulerian 
probability distribution
\begin{equation}
f_{E}^N({\bf u}_1,{\bf x}_1;..;       
 {\bf u}_N,{\bf x}_N;t)=
<\delta({\bf u}_1-{\bf u}({\bf x}_1,t))..
 \delta({\bf u}_N-{\bf u}({\bf x}_N,t))>
\qquad .
\end{equation}
These probability distributions are obtained from the corresponding
Lagrangian (\ref{lg}) ones by integration with respect to ${\bf y}_j$, 
since the fluid flow is incompressible such that the Jacobian (\ref{Jac}) 
equals unity. Furthermore, the Eulerian probability distributions
have to fullfill the following consistency conditions
\begin{eqnarray}
\lefteqn{
\nabla_{{\bf x}_i}
f_E^N({\bf u}_1,{\bf x}_1;..;       
 {\bf u}_N,{\bf x}_N;t)=}
\\&&
[
-\nabla_{{\bf x}'} 
\nabla_{{\bf u}_i}\cdot \int d{\bf u}' {\bf u}'
f_E^{N+1}({\bf u}',{\bf x}',{\bf u}_1,{\bf x}_1;..;       
 {\bf u}_N,{\bf x}_N;t)]_{{\bf x}' ={\bf x}_i}
\qquad i=1,..,N \qquad .
\nonumber 
\end{eqnarray}
These relationships are a consequence of the fact that the spatial dependency
of the probability distribution enters via the velocity field
${\bf u}({\bf x},t)$.

Finally, we mention that it is also
straightforward to define mixed Eulerian and Lagrangian probability
distributions. They can be obtained from the Lagrangian 
probability distribution by integration over the variables ${\bf y}_i$,
for which an Eulerian description is performed:
\begin{eqnarray}
\lefteqn{
f_{E|L}^N({\bf u}_1,{\bf x}_1;..;       
 {\bf u}_m,{\bf x}_m;{\bf u}_{m+1},{\bf x}_{m+1},{\bf y}_{m+1};..;       
 {\bf u}_n,{\bf x}_n,{\bf y}_n;t)=}
\nonumber \\&&
<\delta({\bf u}_1-{\bf u}({\bf x}_1,t))..
 \delta({\bf u}_m-{\bf u}({\bf x}_m,t))
\nonumber \\
&&
\delta[{\bf x}_{m+1}-{\bf X}(t,{\bf y}_{m+1})]
\delta[{\bf u}_{m+1}-{\bf u}({\bf X}(t,{\bf y}_{m+1}),t)]
....
\nonumber \\&&
\delta[{\bf x}_n-{\bf X}(t,{\bf y}_n)]
\delta[{\bf u}_n-{\bf u}({\bf X}(t,{\bf y}_n),t)]>
\nonumber \\&&
=\int d{\bf y}_1...d{\bf y}_m 
f_{L}^N({\bf u}_{1},{\bf x}_{1},{\bf y}_{1};..;       
 {\bf u}_n,{\bf x}_n,{\bf y}_n;t)
\qquad .
\end{eqnarray}

Since the Navier-Stokes equation form a classical field theory
a complete description of the Lagrangian turbulence statistics
is contained in the probability density functional
$P[{\bf x}({\bf y}),{\bf u}({\bf y}),t]$ defined by an average
over functional delta distributions:
\begin{equation}\label{pfunc}
P[{\bf x}({\bf y}),{\bf u}({\bf y}),t]=<{\it D}[{\bf x}({\bf y})-
 {\bf X}(t,{\bf y})]{\it D}[{\bf u}({\bf y})-
 {\bf U}(t,{\bf y})]>
\qquad .
\end{equation}
A probability functional for the field ${\bf X}(t,{\bf y}')$ is
obtained by functional integration with respect to the velocity:
\begin{equation}
G[{\bf x}({\bf y}),t]=\int D{\bf u}({\bf y})
P[{\bf x}({\bf y}),{\bf u}({\bf y}),t]
\qquad .
\end{equation}

The functional Fouriertransform of the quantity (\ref{pfunc}) is the
characteristic functional
\begin{equation}
Z[{\bf \eta}({\bf y}),{\bf k}({\bf y}),t]
=<e^{i \int d{\bf y} [{\bf \eta}({\bf y}) \cdot 
{\bf U}(t,{\bf y})+{\bf k}({\bf y}) \cdot {\bf X}({\bf y},t)]}>
\qquad .
\end{equation}
This is the Lagrangian analog of the functional introduced by 
Hopf \cite{Hopf}.

\section{Hierarchy of Evolution Equations for the N-point
Probability Distribution functions}

In the present section we shall obtain an infinite chain of evolution 
equations for the Lagrangian probability distribution functions
$f^N(\lbrace{\bf u}_j,{\bf x}_j,{\bf y}_j\rbrace;t)$, $N=1,2,...$. 
For the sake of simplicity we consider a fluid in an infinitely
extended space. Therefore, we do not have to take into account
boundary terms. The extention to the case of a fluid in a
bounded area is straightforward.

We start by considering the one-point pdf
$f^1({\bf u},{\bf x},{\bf y};t)$, for which we try to obtain
an evolution equation by calculating the temporal derivative:
\begin{eqnarray}
\lefteqn{
\frac{\partial}{\partial t} f^1({\bf u}_1,{\bf x}_1,{\bf y}_1;t)
=}
\nonumber \\
&&
\frac{\partial}{\partial t}
<\delta[{\bf x}_1-{\bf X}(t,{\bf y}_1)]
\delta[{\bf u}_1-{\bf u}({\bf X}(t,{\bf y}_1),t)]>
\nonumber \\
&& 
=-\nabla_{{\bf x}_1} \cdot <\dot {\bf X}(t,{\bf y}_1) 
\delta[{\bf x}_1-{\bf X}(t,{\bf y}_1)]
\delta[{\bf u}_1-{\bf u}({\bf X}(t,{\bf y}_1),t)]>
\nonumber \\
&&
-\nabla_{{\bf u}_1} \cdot <\dot {\bf u}({\bf X}(t,{\bf y}_1),t) 
\delta[{\bf x}_1-{\bf X}(t,{\bf y}_1)]
\delta[{\bf u}_1-{\bf u}({\bf X}(t,{\bf y}_1),t)]>
\qquad .
\end{eqnarray}
For the first term we can use the relationship 
\begin{eqnarray}
\lefteqn{
<\dot {\bf X}(t,{\bf y}_1) 
\delta[{\bf x}_1-{\bf X}(t,{\bf y}_1)]
\delta[{\bf u}_1-{\bf u}({\bf X}(t,{\bf y}_1),t)]>
=}
\nonumber \\
&&
<{\bf u}({\bf X}(t,{\bf y}_1),t) 
\delta[{\bf x}_1-{\bf X}(t,{\bf y}_1)]
\delta[{\bf u}_1-{\bf u}({\bf X}(t,{\bf y}_1),t)]>
=
\nonumber \\
&&
{\bf u}_1
< 
\delta[{\bf x}_1-{\bf X}(t,{\bf y}_1)]
\delta[{\bf u}_1-{\bf u}({\bf X}(t,{\bf y}_1),t)]>
\qquad .
\end{eqnarray}

In order to evaluate the second term we have to insert the
Lagrangian formulation of the 
Navier-Stokes equation (\ref{lagra}):
\begin{eqnarray}
\lefteqn{
<\dot {\bf u}({\bf X}(t,{\bf y}_1),t) 
\delta[{\bf x}_1-{\bf X}(t,{\bf y}_1)]
\delta[{\bf u}_1-{\bf u}({\bf X}(t,{\bf y}_1),t)]>
=}
\nonumber \\
&&
-\int d{\bf y}'< \Gamma[{\bf X}(t,{\bf y}),{\bf X}(t,{\bf y}')]
:{\bf u}({\bf X}(t,{\bf y}'),t)
:{\bf u}({\bf X}(t,{\bf y}'),t)
\nonumber \\&&
\delta[{\bf x}_1-{\bf X}(t,{\bf y}_1)]
\delta[{\bf u}_1-{\bf u}({\bf X}(t,{\bf y}_1),t)]>
\nonumber \\
&&+ \nu 
\int d{\bf y}' <L({\bf X}(t,{\bf y}),{\bf X}(t,{\bf y}')) 
{\bf u}({\bf X}(t,{\bf y}'),t)
\nonumber \\&&
\delta[{\bf x}_1-{\bf X}(t,{\bf y}_1)]
\delta[{\bf u}_1-{\bf u}({\bf X}(t,{\bf y}_1),t)]>
\qquad .
\end{eqnarray}

The aim is to relate this term in some way to a Lagrangian probability
distribution. This can be achieved by inserting
the identity
\begin{equation}
1=
\int d{\bf u}_2 \int d{\bf x}_2 \delta[{\bf x}_2-{\bf X}(t,{\bf y}')]
\delta[{\bf u}_2-{\bf u}({\bf X}(t,{\bf y}'),t)]
\qquad .
\end{equation}
As a result one obtains:
\begin{eqnarray}
\lefteqn{
<\dot {\bf u}({\bf X}(t,{\bf y}_1),t) 
\delta[{\bf x}_1-{\bf X}(t,{\bf y}_1)]
\delta[{\bf u}_1-{\bf u}({\bf X}(t,{\bf y}_1),t)]>
=}
\nonumber \\
&&
-\int d{\bf u}_2 \int d{\bf x}_2 \int d{\bf y}_2 \Gamma({\bf x}_1,{\bf x}_2)
:{\bf u}_2
:{\bf u}_2 f^2({\bf u}_1,{\bf x}_1,{\bf y}_1;
{\bf u}_2,{\bf x}_2,{\bf y}_2;t)
\nonumber \\&&
+ \nu 
\int d{\bf u}_2 \int d{\bf x}_2
\int d{\bf y}_2 L({\bf x}_1,{\bf x}_2) 
{\bf u}_2 f^2({\bf u}_1,{\bf x}_1,{\bf y}_1;
{\bf u}_2,{\bf x}_2,{\bf y}_2;t)
\qquad .
\end{eqnarray}

We now combine the above formulas to obtain the
evolution equation for the Lagrangian probability distribution
$f^1(\lbrace{\bf u}_1,{\bf x}_1,{\bf y}_1\rbrace;t)$:

\begin{eqnarray}\label{hier11}
\lefteqn{
\frac{\partial }{\partial t} 
f^1({\bf u}_1,{\bf x}_1,{\bf y}_1;t)+
{\bf u}_1 \cdot \nabla_{{\bf x}_1} 
f^1({\bf u}_1,{\bf x}_1,{\bf y}_1;t)
}
\\
&&
=
\nabla_{{\bf u}_1} \cdot  \int d{\bf u}_2 
\int d{\bf x}_2 \int d{\bf y}_2
\Gamma ({\bf x}_1,{\bf x}_2) :{\bf u}_2:{\bf u}_2
f^{2}({\bf u}_1,{\bf x}_1,{\bf y}_1;
{\bf u}_2,{\bf x}_2,{\bf y}_2;t)
\nonumber \\
&&
-
\nu  \nabla_{{\bf u}_1} \cdot  \int d{\bf u}_2 
\int d{\bf x}_2 \int d{\bf y}_2 L({\bf x}_1,{\bf x}_2) {\bf u}_2
f^{2}({\bf u}_1,{\bf x}_1,{\bf y}_1;
{\bf u}_2,{\bf x}_2,{\bf y}_2;t)
\qquad .
\nonumber 
\end{eqnarray}
Due to the nonlocality of the pressure and the viscous term, which relates
the Lagrangian path of a particle under consideration to the paths of
different particles, the evolution equation of the one-particle pdf
is linked to the two-particle pdf. 

It is straightforward to prove that the N-point distribution
function fullfills the evolution equation:
\begin{eqnarray}\label{hier1}
\lefteqn{
\frac{\partial }{\partial t} 
f^N(\lbrace{\bf u}_j   ,{\bf x}_j,{\bf y}_j\rbrace;t)+
\sum_i {\bf u}_i \cdot \nabla_{{\bf x}_i} 
f^N(\lbrace{\bf u}_j,{\bf x}_j,{\bf y}_j\rbrace;t)
}
 \\
&&
=
\sum_i \nabla_{{\bf u}_i} \cdot  \int d{\bf u} 
\int d{\bf x} \int d{\bf y}
\Gamma ({\bf x}_i,{\bf x}) :{\bf u}:{\bf u}
f^{N+1}({\bf u},{\bf x},{\bf y};
\lbrace{\bf u}_j,{\bf x}_j,{\bf y}_j\rbrace;t)
\nonumber \\
&&
-
\nu \sum_i \nabla_{{\bf u}_i} \cdot  \int d{\bf u} 
\int d{\bf x}\int d{\bf y} L({\bf x}_i,{\bf x}) {\bf u}
f^{N+1}({\bf u},{\bf x},{\bf y};
\lbrace{\bf u}_j,{\bf x}_j,{\bf y}_j\rbrace;t)
\qquad .
\nonumber 
\end{eqnarray}
As in the case of the one-point pdf no closed equation for the
N-point pdf is obtained. The evolution equation for the
N-point pdf contains the N+1-point pdf leading to a hierarchy of
equations. 

In the dissipation term we can perform a partial integration
which eliminates the formally defined operator $L({\bf x}_i,{\bf x})$.
Additionally, one can recast the pressure and dissipation terms in a way which
evidences Galilean invariance:
\begin{eqnarray}\label{hier0}
\lefteqn{
\frac{\partial }{\partial t} 
f^N(\lbrace{\bf u}_j,{\bf x}_j,{\bf y}_j\rbrace;t)+
\sum_i {\bf u}_i \cdot \nabla_{{\bf x}_i} 
f^N(\lbrace{\bf u}_j,{\bf x}_j,{\bf y}_j\rbrace;t)
}
 \\
&&
=
\sum_i \nabla_{{\bf u}_i} \cdot  \int d{\bf u} 
\int d{\bf x} \int d{\bf y}
\Gamma ({\bf x}_i,{\bf x}) :({\bf u}-{\bf u}_i):({\bf u}-{\bf u}_i)
\nonumber \\
&&
f^{N+1}({\bf u},{\bf x},{\bf y};
\lbrace{\bf u}_j,{\bf x}_j,{\bf y}_j\rbrace;t)
\nonumber \\
&&
-
\nu \sum_i \nabla_{{\bf u}_i} \cdot  \int d{\bf u} 
\int d{\bf x}\int d{\bf y} \delta({\bf x}_i-{\bf x})
\Delta_{{\bf x}} [{\bf u}-{\bf u}_i]
f^{N+1}({\bf u},{\bf x},{\bf y};\lbrace{\bf u}_j,{\bf x}_j,{\bf y}_j\rbrace;t)
\qquad .
\nonumber
\end{eqnarray}

Let us now introduce the notation
\begin{eqnarray}\label{hier2}
\lefteqn{
\frac{\partial }{\partial t} 
f^N(\lbrace{\bf u}_j,{\bf x}_j,{\bf y}_j\rbrace;t)+
\sum_i {\bf u}_i \cdot \nabla_{{\bf x}_i} 
f^N(\lbrace{\bf u}_j,{\bf x}_j,{\bf y}_j\rbrace;t)
}
 \\
&&
=
-\sum_i \nabla_{{\bf u}_i} \cdot  \int d{\bf u} 
\int d{\bf x} \int d{\bf y}
A({\bf x}_i-{\bf x},{\bf u}_i-{\bf u})
f^{N+1}({\bf u},{\bf x},{\bf y};
\lbrace{\bf u}_j,{\bf x}_j,{\bf y}_j\rbrace;t)
\qquad .
\nonumber 
\end{eqnarray}
Here, $\sigma$ denotes the triple ${\bf u}$, ${\bf x}$, ${\bf y}$.
A is an operator which is related to the acceleration and is defined 
according to:
\begin{equation}
A({\bf x}_i-{\bf x},{\bf u}_i-{\bf u})
=-
\Gamma ({\bf x}_i,{\bf x}) :({\bf u}-{\bf u}_i):({\bf u}-{\bf u}_i)
+
\nu  \delta({\bf x}_i-{\bf x})
\Delta_{{\bf x}} [{\bf u}-{\bf u}_i]
\qquad .
\end{equation}

One may also add a random force field, which is Gaussian as well as
$\delta$-correlated in time. Then a diffusion term of the form
\begin{equation}
\frac{1}{2} \sum_i \sum_j  \nabla_{{\bf u}_i} Q({\bf x}_i,{\bf x}_j)
\nabla_{{\bf u}_j} 
f^{N}(\lbrace \sigma_j\rbrace;t)
\end{equation}
has to be included.

It is important to formulate the following invariance properties
of the hierarchy. Due to
Galilean invariance 
$f^N(\lbrace{\bf u}_i+{\bf c},{\bf x}_i-{\bf c}t,{\bf y}_i\rbrace;t)$ 
solves the hierarchy provided that 
$f^N(\lbrace{\bf u}_i,{\bf x}_i,{\bf y}_i\rbrace;t)$ is a solution
of the hierarchy.
Neglecting the viscous terms, it can be shown that if 
$f^N(\lbrace{\bf u}_i,{\bf x}_i,{\bf y}_i\rbrace;t)$ is a solution, 
then also 
$\lambda^{3(\gamma+\delta)}f^N(\lbrace\lambda^\gamma{\bf u}_i,
\lambda^\delta {\bf x}_i,\lambda^\delta 
{\bf y}_i\rbrace;\lambda t)$ solves the hierachy 
for each value of $\lambda$, provided that 
\begin{equation}
\delta-\gamma=1
\qquad .
\end{equation}
The scale symmetry for an arbitrary value of $\delta$ is obviously broken 
by the viscous term. 

As we have indicated above, the Eulerian N-point pdf can be calculated
by the corresponding Lagrangian pdf by integration with respect to
the initial locations ${\bf y}_i$ of the particles. 
Integrating each equation of the hierarchy (\ref{hier2}) we obtain
a corresponding one for the Eulerian probability distribution function. 
This hierarchy has already been presented by Lundgren \cite{Lundgren}
as well as Ulinich and Lyubimov \cite{Ulinich}.

Let us briefly comment on the question why a whole hierarchy of evolution
equation arises. The mathematical treatment of fluid flow leads to
a field theory. Considering only a finite number of fluid particles,
therefore, yields a description with restricted information. This reduction of
information shows up in the
existence of an unclosed hierarchy of evolution equations for the
joint velocity-position pdfs. A closed evolution equation can only
be expected to arise when one approaches the continuum
description, i.e. when one considers the full probability density
functional defined in eq. (\ref{pfunc}).

\section{Evolution equation for the probability density functional}

This section is devoted to the derivation of a closed evolution equation for
the probability density functional 
$P[{\bf x}({\bf y}),{\bf u}({\bf y}),t] $. Time differentiation yields:
\begin{eqnarray}
\lefteqn{
\frac{d}{dt}P[{\bf x}({\bf y}),{\bf u}({\bf y}),t] =}
\nonumber \\
&&
-\int d{\bf y} \lbrace <
\dot {\bf X}(t,{\bf y}) \cdot \frac{\delta}{\delta {\bf x}({\bf y})}
{\it D}[{\bf x}({\bf y})-
 {\bf X}(t,{\bf y})]{\it D}[{\bf u}({\bf y})-
 {\bf U}(t,{\bf y})]>
\nonumber \\
&&  
+<
\dot {\bf U}(t,{\bf y}) \cdot \frac{\delta}{\delta {\bf u}({\bf y})}
{\it D}[{\bf x}({\bf y})-
 {\bf X}(t,{\bf y})]{\it D}[{\bf u}({\bf y})-
 {\bf U}(t,{\bf y})]> \rbrace
\qquad .
\end{eqnarray}
Using the Lagrangian formulation of the Navier-Stokes equation
(\ref{lagra}) we end up with the following relation:
\begin{eqnarray}\label{hopf}
\lefteqn{
\frac{d}{dt}P[{\bf x}({\bf y}),{\bf u}({\bf y}),t] +
\int d{\bf y}  
{\bf u}({\bf y}) \cdot \frac{\delta}{\delta {\bf x}({\bf y})}
P[{\bf x}({\bf y}),{\bf u}({\bf y}),t] }
\nonumber \\
&&=
\int d{\bf y} \int d{\bf y}' \lbrace
\frac{\delta}{\delta {\bf u}({\bf y})}\cdot 
\Gamma({\bf x}({\bf y}),{\bf x}({\bf y}')):{\bf u}({\bf y}'):{\bf
u}({\bf y}')
\nonumber \\
&& - \nu 
\frac{\delta}{\delta {\bf u}({\bf y})}\cdot
L({\bf x}({\bf y}),{\bf x}({\bf y}')){\bf u}({\bf y}') \rbrace
P[{\bf x}({\bf y}),{\bf u}({\bf y}),t]
\qquad .
\end{eqnarray}
We have arrived at a closed equation determining the evolution
of the probability functional $P[{\bf x}({\bf y}),{\bf u}({\bf
y}),t]$. The N-point probability functions 
$f^N(\lbrace{\bf u}_i,{\bf x}_i,{\bf y}_i\rbrace;t)$ can be obtained
from $P[{\bf x}({\bf y}),{\bf u}({\bf y}),t]$
by functional integration
\begin{eqnarray}\label{proj}
\lefteqn{
f^N(\lbrace{\bf u}_i,{\bf x}_i,{\bf y}_i\rbrace;t)
=\int D{\bf u}({\bf y}) D{\bf x}({\bf y})
\delta[{\bf u}_1-{\bf u}({\bf y}_1)]
\delta[{\bf x}_1-{\bf x}({\bf y}_1)]
}
\nonumber \\
&&.....
\delta[{\bf u}_N-{\bf u}({\bf y}_N)]
\delta[{\bf x}_N-{\bf x}({\bf y}_N)]
P[{\bf x}({\bf y}),{\bf u}({\bf y}),t]
\qquad .
\end{eqnarray}
The hierarchy of evolution equations (\ref{hier2})
for the N-point pdfs is a projection of the functional equation 
(\ref{hopf}) onto the pdf of N fluid particles according to
(\ref{proj}). We mention that, in principle, a projector formalism
should be used to pass from the evolution equation for the probability 
functional (\ref{hopf}) to a closed equation for the projected N-point
probability distribution. 
 
\section{Generalized Fokker-Planck equations}

If we investigate the problem of turbulence by a hierarchy of
evolution equations such as (\ref{hier2}) we need to formulate suitable
closure schemes. In the following we shall present
a formulation of (\ref{hier2}) which seems to be more
suitable for that purpose. It is influenced by the so-called
Lagrangian pdf method \cite{Pope1}, \cite{Pope3} , \cite{Sawford} , 
whose basic idea is to model the acceleration term by a stochastic force. 
Assuming Markovian properties the model pdf obeys a 
Fokker-Planck equation of the form
\begin{eqnarray}
\lefteqn{
\frac{\partial }{\partial t} 
f^N(\lbrace{\bf u}_j,{\bf x}_j,{\bf y}_j\rbrace;t)+
\sum_i {\bf u}_i \cdot \nabla_{{\bf x}_i} 
f^N(\lbrace{\bf u}_j,{\bf x}_j,{\bf y}_j\rbrace;t)
}
\\
&&
=-\sum_i \nabla_{{\bf u}_i} \cdot D^1(\lbrace{\bf u}_j,{\bf x}_j,{\bf
y}_j\rbrace)
\nonumber 
\\
&&
+\sum_{ij} \nabla_{{\bf u}_i}  D^2(\lbrace{\bf u}_j,{\bf x}_j,{\bf
y}_j\rbrace)\nabla_{{\bf u}_j} 
f^N(\lbrace{\bf u}_j,{\bf x}_j,{\bf y}_j\rbrace;t) \qquad .
\nonumber
\end{eqnarray}
The question arises, what functional form of the drift term
$D^1(\lbrace{\bf u}_j,{\bf x}_j,{\bf y}_j\rbrace)$ and
the diffusion matrix 
$D^2(\lbrace{\bf u}_j,{\bf x}_j,{\bf y}_j\rbrace)$ has to be chosen
in order to obtain an accurate model of turbulent flows. Although
the case of a single fluid particle seems to be well-investigated
\cite{Pope3}, the case of several fluid particles has to be
studied in more detail \cite{Novikov}, \cite{Pedrizetti},
\cite{Sawford}.
Interesting models for several fluid particles have been devised by
Pumir et al. \cite{Pumir1}, \cite{Pumir2}. 

Origionally, the random force method dates back to Oboukhov \cite{Oboukhov},
who suggested to use a Fokker-Planck equation without drift
term and a constant diffusion term for the single particle case.

In the following we shall derive a generalization of the
Fokker-Planck equation directly from the hierarchy (\ref{hier2}).
To this end we consider equation (\ref{hier2}) to be a linear inhomogeneous
equation, which can be solved in a straightforward manner
(we consider the case of the (N+1)-point pdf):
\begin{eqnarray}
\lefteqn{
f^{N+1}({\bf u},{\bf x},{\bf y};\lbrace{\bf u}_i,{\bf x}_i,{\bf y}_i\rbrace;t)
=}
\\
&&
e^{-(t-t_0)[{\bf u}\cdot \nabla_{{\bf x}}
+\sum_i {\bf u}_i\cdot \nabla_{{\bf x}_i}]}
f^{N+1}({\bf u},{\bf x},{\bf y};
\lbrace{\bf u}_i,{\bf x}_i,{\bf y}_i\rbrace;t_0)
\nonumber \\
&&
-
\int_{t_0}^t dt'
e^{-(t-t')[{\bf u}\cdot \nabla_{{\bf x}}+
\sum_i {\bf u}_i\cdot \nabla_{{\bf x}_i}]}
\int d\sigma'
\nonumber \\
&&
[\sum_i  
{\bf A}({\bf x}_i-{\bf x}',{\bf u}_i-{\bf u}')
\cdot \nabla_{{\bf u}_i}
+{\bf A}({\bf x}-{\bf x}',{\bf u}-{\bf u}')
\cdot \nabla_{{\bf u}}]
\nonumber \\
&&
f^{N+2}({\bf u},{\bf x},{\bf y};
{\bf u}',{\bf x}',{\bf y}';
\lbrace{\bf u}_j,{\bf x}_j,{\bf y}_j\rbrace;t') \qquad .
\nonumber 
\end{eqnarray}
The first term stems from the initial condition. 

We remind the reader that we have considered the force free case. 
Nonrandom forces ${\bf F}({\bf x},t)$ can be taken into account 
by a different evolution operator 
\begin{equation}
e^{-(t-t'){\bf u}\cdot \nabla_{\bf x} }
\rightarrow
T e^{-(t-t'){\bf u}\cdot \nabla_{\bf x}-\int_t'^t d\tau 
{\bf F}({\bf x},\tau
)\cdot \nabla_{\bf u}} \qquad .
\end{equation}
(Here, T denotes Dysons time ordering operator). Whereas the first
evolution operator makes the replacement 
\begin{equation}\label{zero}
{\bf x} \rightarrow {\bf x}-{\bf u}(t-t')
\end{equation}
the second evolution operator replaces 
\begin{equation}
{\bf x} \rightarrow {\bf X}({\bf x},t-t') \qquad ,
\end{equation}
where ${\bf X}({\bf x},t-t_0)$ is the solution of the
set of differential equations
\begin{eqnarray}
\frac{d}{dt'} {\bf X}({\bf x},t-t') &=& {\bf U}({\bf x},t-t')
\nonumber \\
\frac{d}{dt'} {\bf U}({\bf x},t-t') &=& {\bf F}({\bf X}({\bf x},t-t'),t')
\end{eqnarray}
with the conditions 
\begin{eqnarray}
{\bf X}({\bf x},0) &=& {\bf x} \nonumber \\
{\bf U}({\bf x},0) &=& {\bf u} \qquad .
\end{eqnarray}

Let us make some remarks on the external force. In three dimensional
turbulence the force varies on the so-called integral scale, which
is larger than the scales belonging to the inertial scale. That 
implies that during inertial time scales the relative motion of Lagrangian
particles located within the inertial range  is not influenced by the
external force, i.e. by the mechanism how the turbulence is generated.
Therefore, on time scales belonging to the inertial time scale
the approximation (\ref{zero}) is sufficiently good for our
purposes.   

Now we insert the obtained expression into the acceleration term
of eq. (\ref{hier2}). As a result we arrive at the hierarchy 
\begin{eqnarray}\label{gfpe}
\lefteqn{
\frac{\partial }{\partial t} 
f^N(\lbrace{\bf u}_j,{\bf x}_j,{\bf y}_j\rbrace;t)+
\sum_i {\bf u}_i \cdot \nabla_{{\bf x}_i} 
f^N(\lbrace{\bf u}_j,{\bf x}_j,{\bf y}_j\rbrace;t)
}
\\
&&
=-\sum_i \nabla_{{\bf u}_i} \cdot
\int_{t_0}^t dt' [ {\bf D}^1({\bf x}_i|
\lbrace {\bf u}_i,\tilde
{\bf x}_i,{\bf y}_i\rbrace;t,t')
f^N(\lbrace {\bf u}_i,\tilde{\bf x}_i,{\bf y}_i\rbrace;t')]_{\tilde
{\bf x}_i={\bf X}_i({\bf x}_i,t-t')}
\nonumber \\
&&
+\sum_{ij}
\nabla_{{\bf u}_i} \cdot
\int_{t_0}^t dt' [ {\bf D}^2({\bf x}_i,\tilde{\bf x}_j|
\lbrace{\bf u}_i,\tilde
{\bf x}_i,{\bf y}_i\rbrace;t,t')
\cdot
\nabla_{{\bf u}_j}
f^N(\lbrace{\bf u}_i,\tilde{\bf x}_i,{\bf y}_i\rbrace;t')]_{\tilde
{\bf x}_i={\bf X}_i({\bf x}_i,t-t')}
\nonumber 
\\
&&
-\sum_i \nabla_{{\bf u}_i} \cdot
\int d {\bf \sigma}
{\bf A}({\bf x}_i-{\bf x},{\bf u}_i-{\bf u})
\nonumber 
\\
&& 
f^{N+1}({\bf u},{\bf X}({\bf x},t-t_0),{\bf y};
\lbrace {\bf u}_i, {\bf X}_i({\bf x}_i,t-t_0),{\bf y}_i\rbrace;t=t_0)
\qquad .
\nonumber 
\end{eqnarray}
The last term stems from the initial condition. We want to point out
that, in some sense, the present procedure is analogous to a projection
operator formalism \cite{Mori}. It is well-known that the stochastic
equations contain initial terms after projection. By some more or less
sophisticated arguments, these initial terms are dropped.    

Our result (\ref{gfpe}) takes the form of a generalized Fokker-Planck
equation, where the generalized drift  term
is given by
\begin{eqnarray}\label{drift}
\lefteqn{
D^1({\bf x}_i|\lbrace{\bf u}_i,\tilde
{\bf x}_i,{\bf y}_i\rbrace;t,t'))=}
\nonumber \\
&&
\int d\sigma   \int d\sigma'
{\bf A}({\bf x}_i-{\bf x},{\bf u}_i-{\bf u})
\nonumber \\
&&
\lbrace
[\sum_j {\bf A}(\tilde {\bf x}_j-{\bf x}',{\bf u}_j-{\bf u}')
\cdot \nabla_{{\bf u}_j}+
{\bf A} (\tilde {\bf x}-{\bf x}',{\bf u}-{\bf u}')
\cdot \nabla_{{\bf u}}] 
\nonumber 
\\
&&
p^{N+2}({\bf u}',{\bf x}',{\bf y}';{\bf u},\tilde {\bf x},{\bf y}|
\lbrace{\bf u}_j,\tilde {\bf x}_j,{\bf y}_j \rbrace;t')
\rbrace_{\tilde {\bf x}={\bf X}({\bf x},t-t')}
\qquad .
\end{eqnarray}
The diffusion term takes the form 
\begin{eqnarray}\label{diffusion}
\lefteqn{
D^2({\bf x}_i,\tilde{\bf x}_j|\lbrace{\bf u}_i,\tilde
{\bf x}_i,{\bf y}_i\rbrace;t,t')
=  }
\nonumber \\
&&
\int d\sigma \int d\sigma'
{\bf A}({\bf x}_i-{\bf x},{\bf u}_i-{\bf u}) 
{\bf A}(\tilde {\bf x}_j-{\bf x}',{\bf u}_j-{\bf u}')
\nonumber
\\
&& \times
p^{N+2}({\bf u}',{\bf x}',{\bf y}';{\bf u},\tilde {\bf x},{\bf y}|
\lbrace{\bf u}_j,\tilde {\bf x}_j,{\bf y}_j \rbrace;t')|_{\tilde {\bf
x}=
{\bf X}({\bf x},t-t')}
\qquad .
\nonumber
\end{eqnarray}
A formally closed equation has been obtained by the introduction
of the conditional probability 
distribution
\begin{equation}
f^{N+2}(\sigma'
;\sigma;
\lbrace \sigma_j \rbrace;t)
=p^{N+2}(\sigma'
;\sigma|
\lbrace \sigma_j \rbrace;t)
f^{N}(\lbrace \sigma_j \rbrace;t)
\qquad ,
\end{equation}
($\sigma$ denotes the triple ${\bf u}$, ${\bf x}$, ${\bf y}$.)  
A successful description of Lagrangian turbulence statistics can 
be achieved if this conditional probability distribution can either be
approximated or modeled in a suitable way. Thereby, the fundamental
symmetries, i.e. Galilean and scale invariance (for $\nu=0$), have to
be retained.
Furthermore, incompressibility of the fluid motion should be
conserved. This requirement seems to be the major difficulty, 
since any approximation 
has a consequence for the pressure term. However, only a correct
treatment of the pressure term guarantees incompressibility of the
fluid. 

\section{Summary}

We have formulated a hierachy of evolution equations for
the Lagrangian N-point pdf's in close analogy to
the one for the Eulerian pdf's presented by Lundgren
\cite{Lundgren} and Ulinich and Lyubimov \cite{Ulinich}.  
Due to the pressure and dissipative terms the
N-point probability distributions couple to (N+1)-point
distributions.  The existence of a whole hierarchy of 
evolution equations is due to the fact that a field theory is described 
by a finite number of points. A closed statistical equation
arises when one defines a probability functional. We have
formulated an evolution equation for this functional which
is the Lagrangian analog of Hopf's functional equation.

Furthermore, we have tried to derive the Lagrangian pdf models \cite{Pope1},
which are successful in modeling various aspects of turbulent flows
by diffusion processes. 
Starting from the hierarchy of evolution equations for 
N-point pdfs we arrived at a generalized Fokker-Planck
equation, i.e. a diffusion equation containing memory terms
as well as a term steming from the initial condition.
The generalized drift and diffusion coefficients 
are formally expressed by conditional probability distributions, so
that the problem is not closed. In a following paper we shall 
adress the problem of formulating suitable closure approximations
\cite{Fried}.

\vspace{1cm}

\noindent
Acknowledgement: I gratefully acknowledge financial support from
the Deutsche Forschungsgemeinschaft within the project
({\em Interdisziplin\"are Turbulenzinitiative}). Furthermore, I thank
Joachim Peinke (Oldenburg) and Rainer Friedrich (M\"unchen) for 
interesting discussions.

\end{document}